\documentstyle[12pt,emulateapj,psfig,redo_tabnote]{article}

\submitted{Submitted to The Astrophysical Journal, 1999 March 15.}

\tighten

\lefthead{M. S. Briggs et al.}
\righthead{CGRO Observations of GRB~990123}

\def\g123{GRB~990123}

\begin{document}

\title{Observations of GRB~990123 by the Compton Gamma-Ray Observatory}
\author{
M.~S.~Briggs,\altaffilmark{1},
D.~L.~Band,\altaffilmark{2},
R.~M.~Kippen\altaffilmark{1},
R.~D.~Preece\altaffilmark{1},
C.~Kouveliotou\altaffilmark{3},
J.~van~Paradijs\altaffilmark{1,4},
G.~H.~Share\altaffilmark{5},
R.~J.~Murphy\altaffilmark{5},
S.~M.~Matz\altaffilmark{6},
A.~Connors\altaffilmark{7},
C.~Winkler\altaffilmark{8},
M.~L.~McConnell\altaffilmark{9},
J.~M.~Ryan\altaffilmark{9},
O.~R.~Williams\altaffilmark{8},
C.~A.~Young\altaffilmark{9}
B.~Dingus\altaffilmark{10},
J.~R.~Catelli\altaffilmark{11},
R.~A.~M.~J.~Wijers\altaffilmark{12}
}
\altaffiltext{1}{University of Alabama, Huntsville, AL 35812;
Michael.Briggs, Marc.Kippen, Robert.Preece@msfc.nasa.gov}
\altaffiltext{2}{CASS 0424, University of California,
San Diego, La Jolla, CA
92093; dband@ucsd.edu}
\altaffiltext{3}{Universities Space Research Association,
Huntsville, AL 35800;
Chryssa.Kouveliotou@msfc.nasa.gov}
\altaffiltext{4}{University of Amsterdam, Amsterdam,
Netherlands; jvp@astro.uva.nl}
\altaffiltext{5}{NRL, Washington, DC 20375; share@osse.nrl.navy.mil}
\altaffiltext{6}{Northwestern University, Evanston, IL 60208;
matz@ossenu.astro.nwu.edu}
\altaffiltext{7}{Wellesley College, Wellesley, MA;
aconnors@maria.wellesley.edu}
\altaffiltext{8}{Space Science Department of ESA,
Astrophysics Division, ESTEC, Noordwijk, The Netherlands;
owilliam, cwinkler@estsa2.estec.esa.nl}
\altaffiltext{9}{University of New Hampshire, Durham, NH 03824;
james.ryan@unh.edu, ayoung@comptel.sr.unh.edu}
\altaffiltext{10}{University of Utah, Salt Lake City, UT 84102;
dingus@mail.physics.utah.edu}
\altaffiltext{11}{Goddard Space Flight Center, Greenbelt, MD;
jrc@egret.gsfc.nasa.gov}
\altaffiltext{12}{Dept. of Physics and Astronomy,
SUNY, Stony Brook, NY 11794-3800; rwijers@astro.sunysb.edu}
\begin{abstract}
\g123\ was the first burst from which simultaneous optical, X-ray and
gamma-ray emission was detected; its afterglow has been followed by
an extensive set of radio, optical and X-ray observations.
We have
studied the gamma-ray burst itself as observed by the {\it CGRO}
detectors.
We find that gamma-ray fluxes are not correlated with
the simultaneous optical observations, and the gamma-ray spectra
cannot be extrapolated simply to the optical fluxes.
The burst is well fit by the standard four-parameter GRB function, with the
exception that excess emission compared to this function is observed
below $\sim 15$~keV during some time intervals.
The burst is
characterized by the typical hard-to-soft and hardness-intensity correlation
spectral evolution patterns.
The energy of the peak of the $\nu f_\nu$ spectrum,
$E_{\rm p}$, reaches an unusually high value during the first intensity spike,
$1470 \pm 110$ keV, and then falls to  $\sim$300~keV
during the tail of the burst.
The high-energy spectrum above
$\sim1$~MeV is consistent with a power law with a photon index of about $-3$.
By fluence, GRB~990123 is brighter than all but 0.4\% of
the GRBs observed with BATSE,
clearly placing it on the $-3/2$ power-law portion of the intensity
distribution.
However, the redshift measured for the afterglow is inconsistent with
the Euclidean interpretation of the $-3/2$ power-law.
Using the redshift value of $\geq 1.61$ and assuming isotropic emission,
the gamma-ray fluence exceeds $10^{54}$ ergs.
\end{abstract}
\section{INTRODUCTION}
\g123\ was the first gamma-ray burst to be simultaneously detected in
the optical band.  The
Robotic Optical Transient Search Experiment
(ROTSE) detected optical emission  during the burst and also
immediately after gamma-ray emission was no longer
detectable (Akerlof et al. 1999a,b).
Observations of prompt X- and gamma-rays were made by several spacecraft,
including
all four instruments on the {\it Compton Gamma-Ray Observatory},
and the Wide Field Camera on {\it Beppo-SAX} (Feroci et al. 1999).
The burst's afterglow was detected and monitored at X-ray (Heise et
al. 1999), optical (Odewahn et al. 1999) and radio (Frail et al.
1999) energies, resulting in the rapid localization of the burst
(Piro et al. 1999; Heise et al. 1999), the determination of the redshift
($z\ge1.61$---Kelson et al. 1999; Hjorth et al. 1999), and the
characterization of the afterglow's spectrum and evolution (Galama
et al. 1999, Fruchter et al. 1999b, Kulkarni et al. 1999, Sari \& Piran 1999b).
This extensive set of observations permits an
analysis of the burst's relation to its afterglow. Here we present
an analysis of the observations obtained with the instruments on the
{\it Compton Gamma-Ray Observatory} (CGRO). Our goals are to: a)~relate
the X-ray and gamma-ray spectra to the optical flux;
b)~to characterize the burst, comparing it to other bursts, and
c)~to discuss the implications of this event.

According to the currently favored class of theoretical models,
the observed burst and
subsequent afterglow radiation are synchrotron emission by
non-thermal electrons accelerated by strong shocks in relativistic
outflows (Rees \& M\'esz\'aros 1992; Galama et al. 1998; see Piran
1999 for a review). In this model, the high-energy emission of the
burst itself is radiated by ``internal shocks'' which result from
the collisions between regions with different velocities (Lorentz
factors) within the relativistic outflow, while the lower energy
(radio, optical and X-ray) afterglow is thought to be emitted by an ``external
shock'' where the outflow plows into the surrounding medium.
M\'esz\'aros \& Rees (1997) and Sari \& Piran (1999a)
pointed out that when the external shock first forms, a reverse
shock propagates back into the relativistic outflow; the shocked
region behind the reverse shock is denser and somewhat cooler than
the region behind the external shock, and radiates in the optical
band. The optical emission from the reverse shock will be visible
during or soon after the high-energy burst produced by the internal
shocks.

Multi-wavelength observations during and immediately after the
burst can test the current theoretical model's predictions by
detecting the transitions between the emissions from the different
regions.  Soft X-ray
tails, possibly the beginning of the X-ray afterglow,
are sometimes observed after higher energy emission has ended
(e.g., Yoshida et al. 1989; Sazonov et al. 1998). The X-ray
emission at the end of GRB~970228 (Costa et al. 1997) and
GRB~980329 (in't Zand et al. 1998) is consistent with an
extrapolation of the X-ray afterglow decay observed many hours
after the burst, suggesting that the burst merges smoothly into the
afterglow. GRANAT/SIGMA observed a $t^{-0.7\pm0.03}$ power law
decay of the 35--300~keV  light curve for 1000~s following GRB~920723
(after which the extrapolated flux would have been
undetectable---Burenin et al. 1999); the power law begins
immediately after the burst appears to end, when there is an abrupt
change in the gamma-ray spectral index from 0 to $-1$ (i.e., to
$N_E\propto E^{-1}$).
In prompt pointed observations OSSE detected significant persistent
emission above 50 keV ($\ge 300$ keV in some cases) following the main
burst in four of five events in a fluence selected sample (Matz et al.
1999). The emission was detectable for $10^2$--$10^3$~s with decays
consistent with the power-law declines seen in the afterglows at lower
energies.  There was no evidence for a significant gap between the end
of the burst and the beginning of the persistent emission.
BATSE observations of GRB~980923 show a power-law decay in time
of the emission above
25~keV between about 40 and 400~s after the beginning of the burst---this
decay appears to be a higher-energy analog of the x-ray afterglows seen
in other events (Giblin et al. 1999).
%
On the other hand, the X-ray light curve of GRB~780506
observed by {\it HEAO-1} shows a period of a few minutes without
detected flux after the burst,
followed by renewed emission,
suggesting a gap between X-ray
emission from the burst and from the afterglow (Connors \&
Hueter 1998).

The ROTSE detection of optical flux simultaneous with the gamma-ray
emission and during the $\sim10$~minutes following the burst
(Akerlof et al. 1999a,b) can probe the issue of which physical
regions radiate when. These optical observations consisted of three
5~s exposures beginning 22.2, 47.4 and 72.7~s after the trigger of
{\it CGRO}'s Burst and Transient Source Experiment (BATSE)
at 35216.121~s UT on 1999 January 23  (see Fig.~1),
and three 75~s exposures
beginning 281.4, 446.7 and 612.0~s after the trigger.
Gamma-ray emission was detected for about 100~s, so the last three optical
detections by ROTSE occurred after the apparent end of the burst phase of the
GRB.
ROTSE uses an unfiltered CCD,
but an equivalent V band magnitude is reported; we will use these
magnitudes without attempting any color corrections, etc. The
optical flux increased from the first to the second exposure
($m_v=11.82$ to $m_v=8.95$), and then decayed from exposure to
exposure ($m_v=10.08$, 13.22, 14.00 and 14.53).

All four {\it CGRO} instruments detected this burst (however, the
spark chamber of EGRET was not operating); in fact, ROTSE
responded to a preliminary BATSE position. Here we describe the
spectral evolution of the burst, and compare its spectrum with the
optical flux measurements.  In \S 2 we describe our analysis of
the {\it CGRO} observations of \g123,
while in \S 3 we discuss the implications of this analysis for
burst models.
\section{DATA ANALYSIS}
\subsection{Multi-instrument analysis}
The {\it CGRO} instruments observed \g123\ at different energies
and time scales.
Here we present our procedures and some general results; more detailed
discussion of the data of each instrument appears in the following
subsections.

Figure~1 shows the light curve in different energy
bands accumulated by BATSE and COMPTEL, the two instruments
with the best combination of temporal and spectral
resolution for this purpose.
The light curves show that
the low-energy emission persists longer than high-energy
emission---typical ``hard-to-soft'' spectral evolution.
Particularly striking is the virtual absence of high-energy emission
45--90~s after the burst trigger.
In the highest energy band shown, 4 to 8 MeV, emission is  seen from
the first spike but appears to be absent or much reduced
in the second, providing evidence
that the first spike is harder than the second.
The first spike also appears narrower in the 4 to 8 MeV band, even compared to
the 2 to 4 MeV band.

\begin{figure*}[tbp!]
\hspace{4mm}
\psfig{file=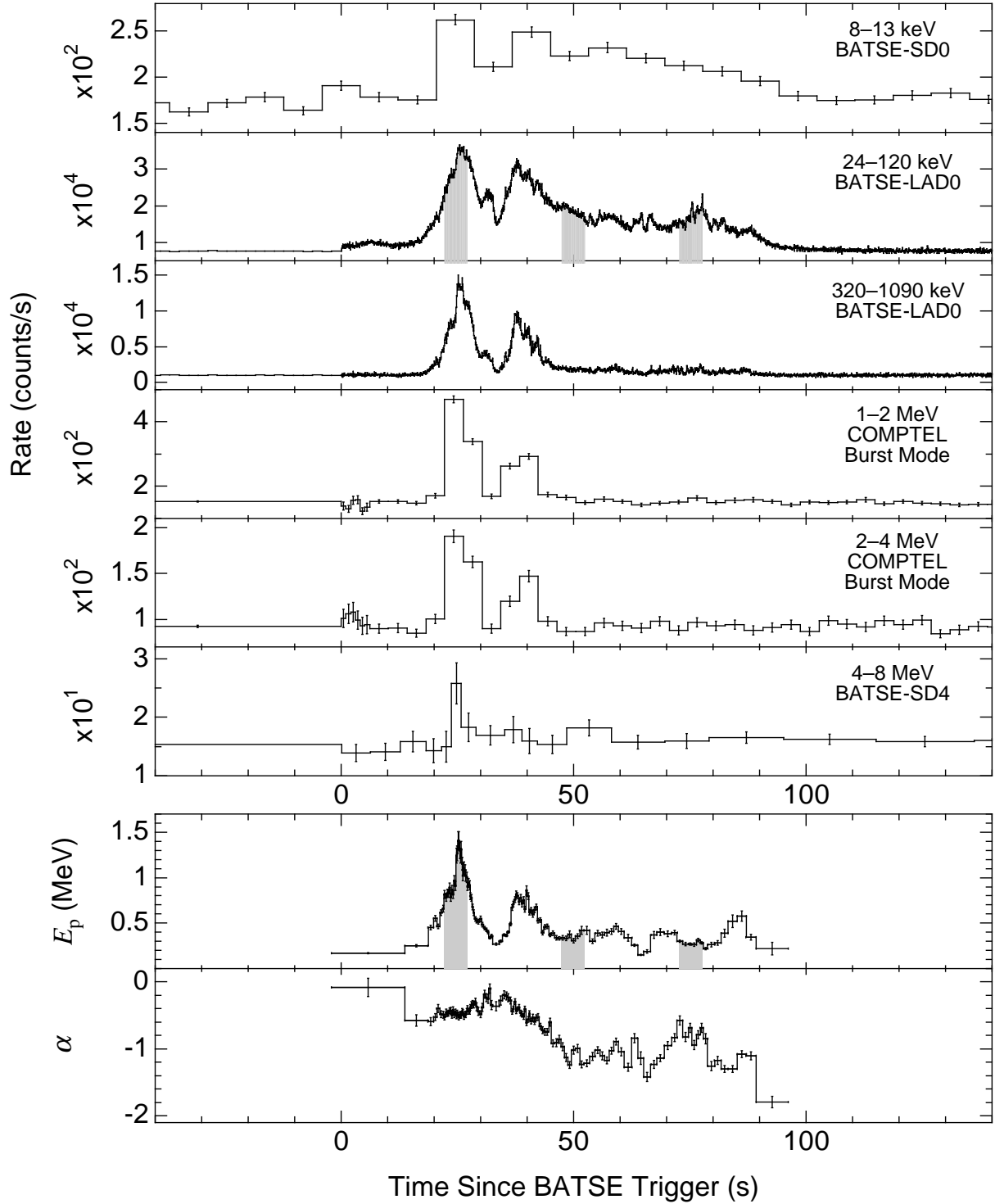,width=165mm,%
bbllx=85bp,bblly=150bp,bburx=510bp,bbury=680bp,clip=}
%
\caption{Top six graphs: Time history of \g123\ using data from  {\it
CGRO} detectors. The intervals over which the first three ROTSE
observations were accumulated are indicated with shaded bands.
Lower two graphs: The time evolution of
$E_{\rm p}$, the energy of the peak of $\nu f_\nu$, and of
$\alpha$, the low-energy photon
index in the GRB function.
These parameters were obtained from
GRB function fits to MER 16 channel spectra
from BATSE LADs 0 and 4.
The horizontal bars on the points  show the time over which the spectrum was
fit, and the vertical bars show the $1\sigma$ uncertainty in the
values.}
\end{figure*}

Fitting models to spectra provides the most accurate
characterization of spectral evolution, although fits are possible
only for spectra with a sufficiently large signal-to-noise ratio
(S/N).  For fitting the BATSE data we use the
``GRB'' function (Band et al. 1993), which consists of a low-energy
power law with an exponential cutoff, $N_E \propto E^\alpha
\exp[-E(2+\alpha)/E_{\rm p}]$, which merges smoothly with a high-energy
power law, $N_E\propto E^\beta$. The ``hardness'' of this spectrum
can be characterized by the energy $E_{\rm p}$ of the peak of
$E^2N_E\propto \nu f_\nu$ for the low-energy component (if
$\beta<-2$, as is the case for GRB~990123,
then $E_{\rm p}$ is the peak energy for the entire spectrum).
Most of the spectral curvature is found in the observations made with
BATSE below about 1 MeV; the spectra above $\sim$ 1~MeV from the other
instruments
are consistent with a simple power-law, which
we identify as the high-energy power law $E^\beta$ of the GRB function.

The model fitting is done using the standard forward-folding technique
(see review by Briggs 1996).  Briefly,
for each instrument we assume a photon model (`GRB' for BATSE, power law
for the other instruments) and convolve that photon spectrum through a detector
model to obtain a model count spectrum.  The model count spectrum is compared
to the observed count spectrum with a goodness-of-fit
statistic, and the photon
model parameters are optimized so as to minimize the statistic.
In most cases $\chi^2$ is used as the goodness-of-fit statistic.
This procedure is model dependent.   Presenting a comparison of the data
and model in count space is the best way to show the quality of the fit.
However, it
is difficult to show results from all four instruments on one
count-rate figure because of the
widely differing responses of the instruments.
We therefore present photon spectra:
the photon ``datapoints'' are calculated by scaling the observed count
rate in a given channel by the ratio of the photon to count model
rate for that channel; this ratio and therefore the photon
datapoints are model dependent.

An additional complication is that
the differing time boundaries of the spectral data from the
{\it CGRO} instruments preclude forming a multi-instrument spectrum
for exactly the same time interval.
We have therefore
selected time intervals which are as similar as possible and which
include the two spikes of the burst (see Table~1).
At high energies the flux of the spikes is much higher than the flux of
non-spike intervals, consequently as long as each spectrum includes the spikes,
differences in the time boundaries are relatively inconsequential, excepting
that the spectra must be normalized to a common time interval encompassing the
spikes.
Because of the small differences expected from the differing time intervals
and because of inter-detector calibration uncertainties, we have not
attempted simultaneously fitting the data from different instruments with
the same photon model.

\begin{deluxetable}{c c r r r r r r c}
\tablefontsize{\scriptsize}
\tablecolumns{9}
\tablewidth{0pc}
\tablecaption{Multi-Instrument Spectral Analysis}
\tablehead{
  &
  &
\multicolumn{3}{c}{Integration Times}  &
\multicolumn{2}{c}{Energy Range\tablenotemark{\S}} &
              &
\colhead{Model Normalization\tablenotemark{\ddagger}}
   \nl
\colhead{instrument} &
\colhead{detector} &
\colhead{start\tablenotemark{\dagger}} &
\colhead{end\tablenotemark{\dagger}} &
\colhead{length}  &
\colhead{low} &
\colhead{high}  &
\colhead{$\beta$}  &
\colhead{at 1 MeV} \nl
  &
  &
\multicolumn{1}{c}{(s)}  &
\multicolumn{1}{c}{(s)}  &
\multicolumn{1}{c}{(s)}  &
\multicolumn{1}{c}{(MeV)}  &
\multicolumn{1}{c}{(MeV)}  &
                 &
\colhead{($\gamma$ cm$^{-2}$ s$^{-1}$ MeV$^{-1}$)}
}
\startdata
BATSE  &  LAD~0       &  12.288 & 45.056  &  32.768  &  0.033 &  1.80  &
$-3.11 \pm 0.07$   &  $1.93 \pm 0.02$  \nl
BATSE  & SD~0 disc.   &  12.288 & 45.056  &  32.768  &  0.008  &  0.013 &
 \nl
BATSE  & SD~1 disc.   &  12.288 & 45.056  &  32.768  &  0.015  &  0.030 &
 \nl
BATSE  & SD~4         &  12.736 & 44.928  &  32.192  &  0.320 &  25. &
 \nl
OSSE   & Detector 2   &  12.288 & 45.056  &  32.768  &   1.0    &  10. &
$-2.82 \pm 0.16$   &  $1.68 \pm 0.23$ \nl
COMPTEL & BSA         &  14.199 & 46.398    &  32.199  &  1.0   &  8.5  &
$-2.78 \pm 0.16$   &  $1.49 \pm 0.10$ \nl
COMPTEL &  telescope  &  13.331 & 46.099  &  32.768  &  0.75  &  30.  &
$-3.33 \pm 0.26$    &  $2.0 \pm 0.26$  \nl
EGRET   &  TASC      & $-$0.057 & 64.503  &  64.560  &  0.97  &  250. &
$-2.71 \pm 0.08$   &  $1.09 \pm 0.07$  \nl
\enddata
\tablenotetext{\dagger}{\footnotesize 
Relative to BATSE trigger time of 35216.121~s.}
\tablenotetext{\S}{\footnotesize
Energy range of the data used in the fit.}
\tablenotetext{\ddagger}{\footnotesize
Systematics dominate the statistical errors.}
\end{deluxetable}

The resulting photon spectra are shown in Figure~2.
The overlap between instruments is at the upper-end of the break region and
in the high-energy power law regime.
The overall agreement between the four instruments and the six
detector types is very good.
Data from the BATSE detectors, LAD~0 and SD~4, overlap from 0.32 to 1.8~MeV
and are in excellent agreement except for the highest energy LAD point
at 1.1 to 1.8~MeV.
The most discordant points are
the highest energy point from BATSE LAD~0 and
the two lowest energy points from the EGRET TASC.
Because of the very small statistical uncertainties on these
three points, systematics
in the calibrations probably dominate.
The values of the high-energy spectral index $\beta$
obtained by the instruments are largely consistent (Table~1).
There are larger differences between the model normalizations (Table~1).
Differences in the normalizations at the level of $\sim 10\%$ are expected
because of uncertainties in the effective areas.
The assumption of a simple power-law for
the OSSE, COMPTEL and EGRET fits, despite the presence
of some curvature in the lowest energy data of these instruments, also
contributes to the discrepancies.

\begin{figure*}[tbp!]
\hspace{20mm}
\psfig{file=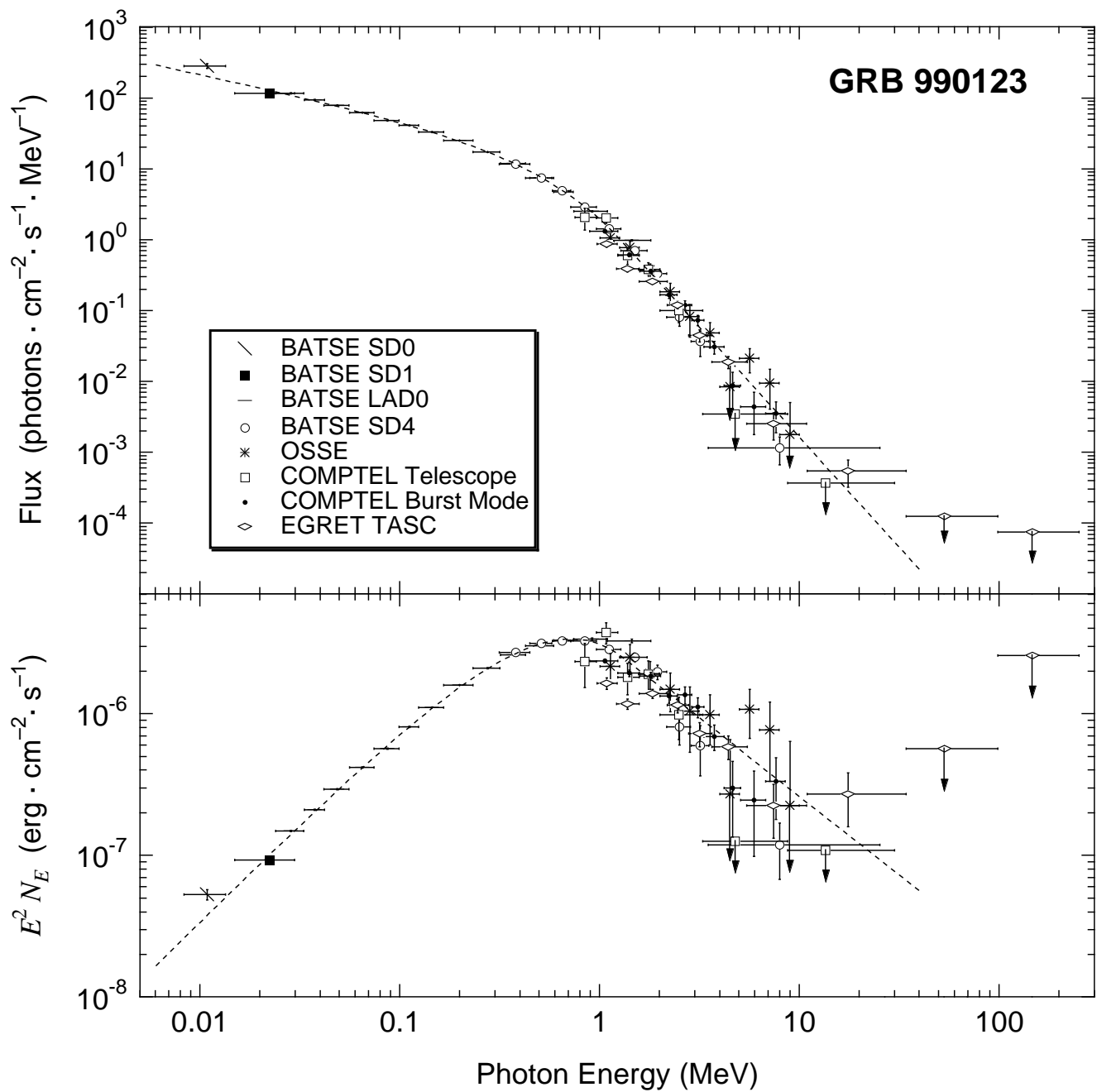,width=180mm,%
bbllx=70bp,bblly=250bp,bburx=510bp,bbury=685bp,clip=}
\caption{Deconvolved spectra from the  {\it CGRO} detectors, shown
both as photon flux $N_E$ and in $E^2 N_E= \nu f_\nu$ units.
The spectra have been rebinned into wider bins for clarity.
Each spectra is calculated using the actual
accumulation times (Table~1), except
for the EGRET TASC spectrum, which uses a shorter time interval during which the
emission was intense (see text)
}
\end{figure*}

\subsection{BATSE}
The different BATSE datatypes provide varying temporal and spectral
resolution.  Each of the eight BATSE modules
contains a Large Area Detector (LAD), a relatively thin 2000 cm$^2$ NaI
crystal, and a Spectroscopy Detector, a 12.7~cm diameter by 7.6~cm thick NaI
crystal.
The large area  of the LADs permit high temporal resolution
analyses, at limited
spectral resolution, while the Spectroscopy Detector (SD) permit higher
spectral resolution studies. The LADs are all operated with the same detector
settings but the SDs are run at different gain settings and
therefore cover a variety of energy ranges. Spectra are accumulated
more frequently for the modules with higher count rates.
Fortunately, the SD with the greatest count rate, SD~0, was in a
high gain state, resulting in  spectra from $\sim25$ to
1750~keV (``SHERB'' spectra), as well as a calibrated discriminator
channel (``DISCSP1'') between 8 and 13~keV. SD~0 was at a burst
angle of 35.2$^\circ$. The
second ``rank'' detector, SD~4, was at a burst angle of
53.8$^\circ$; with a low gain setting, SD~4 provided spectra
between 320~keV and 25~MeV. A discriminator channel between
15 and 27~keV was obtained from the third rank detector SD~1 at
an angle of 58.6$^\circ$.

The LADs provide data from about 33 to 1800~keV in 16 channels.
For Fig.~2, we use the CONT data from LAD~0, which provides
2.048~s resolution.
The MER datatype from the
LADs provide rates in the same 16
channels every 16~ms for the first 32.768~s of the burst, followed
by 64~ms resolution for an additional 131~s.
For GRB~990123, the MER data is summed from LADs~0 and~4, which had
angles to the burst of $27.5^\circ$ and $46.0^\circ$, respectively.
(For GRB~990123,
high time resolution
data from the energy channel from 230 to 320 keV is missing due
to a telemetry gap; however,
all channels are available at 2.048~s resolution via the CONT datatype).
These MER rates show the burst's temporal
morphology (see Figure~1), and are particularly useful for studying
spectral evolution.

Figure~1 (lower panels)
shows the evolution of $E_{\rm p}$ using fits to 16 channel
MER spectra from LADs~0 and 4 rebinned in time to provide  S/N of at least 100.
In order to improve the reliability of the fits, and because there is
little evidence for temporal variations in
$\beta$, the GRB function was used with $\beta$ fixed at $-3.11$.
This value of $\beta$ was obtained from the joint fit to the BATSE data
shown in Fig.~2 and is consistent with the values obtained from the other
instruments (see Table~1).
As can be seen, $E_{\rm p}$
increases by a large factor every time there is a spike in the
light curve, as is typical of ``hardness-intensity'' spectral evolution.
Additionally, the maximum $E_{\rm p}$ is greater in the first spike than in
the second,
$E_{\rm p}$ decreases more rapidly than the count rate and has an overall
decreasing trend, behaviors which are typical
of ``hard-to-soft'' evolution (Ford et al. 1995).
The small $E_{\rm p}$ maximum for the second spike is consistent with the
absence of that spike in the 4 to 8 MeV lightcurve (Fig.~1).
Two intervals during the first spike have $E_{\rm p}$ values
of $1470 \pm 110$ keV.
Such values are exceptional---only 3 bursts of the 156
studied by Preece et al. (1999) have spectra with $E_{\rm p}$ values above
1000 keV.

To investigate the burst spectrum over the broadest energy range
possible, we extend the LAD spectra by also fitting the SD data.
The high-energy resolution SHERB data
can be fit
satisfactorily by the GRB function discussed above; the fits are
consistent with the fits to other data types.
SD~4 provides detections of burst flux to at least the 4.0 to 8.0~MeV band
(Fig.~1).
For the multi-instrument fit shown in Fig.~2, the
BATSE data from LAD~0, SD~4 and SD discriminators 0 and 1 are jointly
fit to a common photon model, allowing a small ``float'' in
normalization between the detectors (for the discriminators, the relative
normalizations are found using the high-resolution data from the same
detectors).
The relative normalization between LAD~0 and SD~4 is consistent with
unity: $1.01 \pm 0.01$.   The relative normalization differences
between the SDs are 10\%, which is a typical value.
The result of the fit to this 32~s interval
is $E_{\rm p}=720 \pm 10$ keV, $\alpha = -0.60 \pm 0.01$ and
$\beta=-3.11 \pm 0.07$.
At 1~MeV, the fitted flux is
$1.93 \pm 0.02$~photons cm$^{-2}$ s$^{-1}$ MeV$^{-1}$.


To test for the presence of an X-ray excess in GRB~990123 (as is seen
in some other bursts---Preece et al. 1996),
we fit the SHERB spectrum and the discriminator channel accumulated by
SD~0 over various segments of the light curve,
including each of the first
three ROTSE observations.
We used the GRB function and fit all data points.
For SD~0 the discriminator channel covers
8--13~keV, and the SHERB data 26--1760~keV. A significant
soft excess above the fit of the GRB function is present in both of the
major emission spikes (i.e., between 17.280 and 44.928~s after the
BATSE trigger), including the first ROTSE observation, but is not
evident in tail of the burst (i.e., after 44.928~s after the
trigger), including the second and third ROTSE observations. This soft
excess is also not present during the weak emission before the
first major emission spike (i.e., up to 14.464~s after the trigger),
and surprisingly, during a weak secondary peak on the declining
edge of the first emission spike (29.696--34.304~s after the
trigger). To verify the presence of the soft excess in the first
ROTSE observation, we added the spectrum accumulated by SD~1 which was at
a lower gain and therefore covered a higher energy range. In
particular, the SD~1 discriminator channel partially covers the gap
between SD~0's discriminator channel and SHERB spectra. As shown by
Figure~3, the joint fit is satisfactory, and the X-ray excess is
weakly evident in the SD~1 discriminator channel.

\begin{figure*}[tbp!]
\begin{center}
\mbox{
\psfig{file=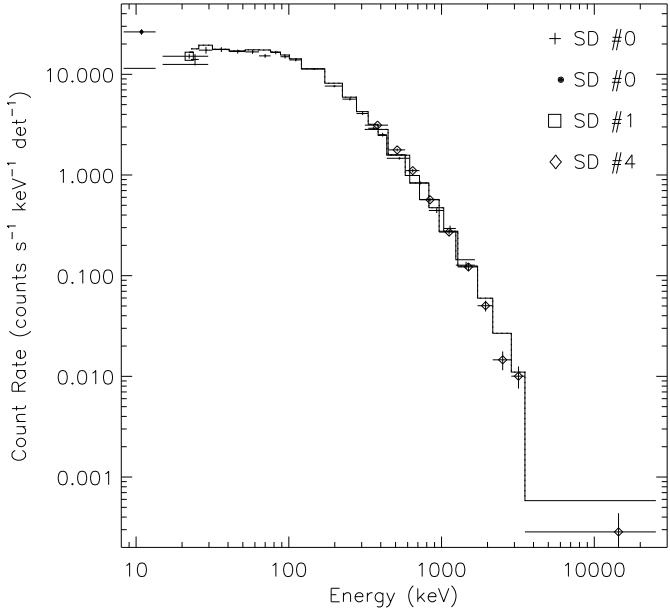,width=90mm,%
bbllx=20bp,bblly=180bp,bburx=540bp,bbury=670bp,clip=}
\hspace{4mm}
\begin{minipage}[b]{85mm}
\caption{Fit to SD count spectra for the time interval of the
first ROTSE observation.
The data points are from the SHERB data of SD~0 and SD~4, and from the
low-energy discriminators of SD~0 and SD~1, as indicated.
The model photon spectrum, the GRB function,
has been folded
through the detector response, resulting in the model count
spectra (histograms).
Each detector, including each discriminator, has a separate count model
and therefore a separate histogram.
The histogram for each discriminator appears as a horizontal line.
Note that the lowest energy
discriminator point clearly lies above the GRB model (horizontal line).
The spectral parameters of the model are
$E_{\rm p}=1170 \pm 30$ keV, $\alpha = -0.63 \pm 0.02$ and
$\beta = -4.1 \pm 0.4$.
}
\protect \vspace{4.5mm}
\end{minipage}
}
\end{center}
\end{figure*}

Integrating spectral fits over the energy range 20--2000~keV
provides a measure of the intensity of the high-energy emission.
During the three ROTSE observations the photon fluxes are 30.98,
8.17 and 7.74 photons s$^{-1}$ cm$^{-2}$ while the energy fluxes
are $2.14\times 10^{-5}$, $1.68\times 10^{-6}$ and $1.82\times
10^{-6}$ erg s$^{-1}$ cm$^{-2}$; the ratios of these fluxes show
that the average photon energy decreased from the first to the
second and third observations. Integrating the GRB function fit to the
SHERB spectrum accumulated over the entire burst gives a $>20$~keV
fluence of $3.0\times10^{-4}$ erg cm$^{-2}$.  The GRBM
of {\it Beppo-SAX}
reported a 40--700~keV fluence of $3.5(\pm0.4)\times10^{-4}$ erg
cm$^{-2}$ (Feroci et al. 1999). The $>20$~keV fluence
$5.09(\pm0.02)\times10^{-4}$ erg cm$^{-2}$ reported by Kippen et
al. (1999) was obtained using BATSE 4-channel discriminator data.
This technique is less accurate, particularly for hard bursts like
GRB~990123, because assumptions about the spectrum must be made in
the deconvolution process.
These diverse fluence values gives a sense of the
consequences of differences in the inter-detector calibrations and
analysis methods.

The low-energy SD discriminator channel (DISCSP datatype) extends
our spectrum down to 8~keV for SD~0. The WFC on {\it Beppo-SAX}
covers the energy range 1.5--26~keV. Feroci et al. (1999) report
that the WFC flux peaked $\sim40$~s after the high-energy peak,
which would place the WFC peak between the second and third ROTSE
observations. The WFC peak intensity of 3.4~Crab corresponds to
$\sim17$ photons s$^{-1}$ cm$^{-2}$ in their energy band. We fit the
DISCSP and SHERB data for SD~0 over $\sim9$~s around the time of the
WFC peak; only the low-energy part of the GRB function was
necessary to fit the spectrum. The DISCSP data point was
$\sim$40\% higher ($\sim1.5\sigma$) than the model fit,
suggesting that there is at most a moderate soft excess.
Integrating the model fit over the WFC energy range gives a flux of
$\sim16$ photons s$^{-1}$ cm$^{-2}$, which is in acceptable agreement
with the WFC flux, given the uncertain inter-detector calibration.
\subsection{OSSE}
The Oriented Scintillation Spectrometer Experiment (OSSE) consists
of 4 collimated ($3.8^\circ\times11.4^\circ$) NaI detectors
covering the ~50~keV to ~250~MeV energy range. The burst occurred
$\sim20^\circ$ and $\sim60^\circ$
outside of the detector's narrow
and broad aperture dimensions, respectively.
However, because the burst was
intense and hard, radiation penetrated the shielding, permitting a
$>1.0$~MeV spectrum to be accumulated. In addition, a high time
resolution (16~ms) light curve was recorded above $\sim$0.13 MeV by the
NaI shields for the first 60~s of the burst.
The time structure is complex with variations observed as short as $\sim$
16~ms.
%
Light curves can be formed in different energy bands using the
photons which penetrated the shields into the detectors. These
light curves show that most of the emission above a few hundred keV
occurred during the two intensity spikes at the beginning of the
burst; this is consistent light curves of BATSE and COMPTEL
(Fig.~1). The spectrum appears to extend to the 3--10~MeV
band.

The spectrum of these photons can be analyzed, even though
they did not originate in the detectors' field-of-view and were
attenuated by the shields, because an appropriate response model
was created and tested on solar flares at comparable off-axis
angles.
The spectrum from Detector~2  suffered the least
scattering and attenuation by material in the spacecraft. The
background spectrum uses data from 115~s before and after the
burst. A power law model was fit to the data, resulting in a 1~MeV
flux of 1.68$\pm0.23$ photons cm$^{-2}$ s$^{-1}$ MeV$^{-1}$ and a
photon index $\beta$ of $-2.82\pm0.16$; this fit was used to create the
photon spectrum shown on Figure~2. Integrating this spectrum over
the 32.768~s of most of the high-energy emission results in a fluence
of $(0.91\pm0.27)\times 10^{-4}$  erg cm$^{-2}$ from 1 to 10~MeV.

\subsection{COMPTEL}

COMPTEL  detected with high significance the two main spikes of GRB~990123 in
both its imaging telescope (``double scatter''; 0.75--30 MeV)  and
non-imaging burst-spectroscopy (``single scatter''; 0.3--1.5 MeV and
0.6--10 MeV) modes.
The burst-spectroscopy or BSA mode relies upon two NaI detectors in the
bottom layer of the telescope
(see Sch\"onfelder et al. 1993 for instrument
details).

In imaging telescope mode, COMPTEL provides detailed information on
individual time-tagged photons with ${{1}\over{8}}$ ms time
resolution.  However, for GRB~990123 these telescope data
suffered from two limitations:
1) the telescope-mode effective area was at best a few cm$^2$
as GRB~990123 occurred nearly 60$^\circ$ from COMPTEL's pointing
direction; and
2) because of the event's high intensity, even at this low effective
area a total of about 16\% of this burst's livetime was lost due to telemetry
limits (a maximum throughput of about 20 events per second).

In contrast, the effective area in burst spectroscopy mode
was roughly two orders of
magnitude greater than in telescope mode, and the deadtime was negligible.
However, the spectral accumulation
time was set at 4
seconds.
Therefore the
COMPTEL light-curves displayed in Fig.~1 are from the burst
spectroscopy mode.  The signal was negligible above 4 MeV.
For both telescope and burst spectroscopy modes, the background
appeared stable so results depended little (a few \%) on the choice of
background integration time.

Spectra from the two modes were handled differently.  For the
0.75--30~MeV telescope data, one selects only events falling within a
certain angle of the source position (``angular resolution measure'', or
A.R.M.; Sch\"onfelder et al. 1993), both reducing background and
creating a nearly diagonal response (e.g. Kippen et al. 1998). For
the telescope spectrum displayed in Fig.~2, a 20$^\circ$
A.R.M. limit was used.  The 32.768~s integration interval (see Table~1)
was chosen both to cover all the significant gamma-ray
emission and to allow the best live-time calculations.  The
background was taken from 131 seconds of data 15 orbits prior to the
burst (see Kippen et al. 1998).
These data were fit via the
forward-folding technique assuming a simple power-law.
Because of the few counts per bin and the background level,
the goodness-of-fit statistic used was the
background-marginalized Poisson statistic of Loredo (1992).
The statistic indicates that this fit is good; however, the data are also
consistent with a break at or below 1 MeV (as indicated by the fit to the
BATSE data) and also with a break at or above 6 MeV.
The best-fit was
$2.0 \pm 0.26 (E/1 {\rm MeV})^{-3.33 \pm 0.26}$
photons cm$^{-2}$ s$^{-1}$ MeV$^{-1}$,
giving a total
fluence of $(1.0 \pm 0.3) \times 10^{-4}$ ergs cm$^{-2}$
(0.75--30 MeV).
Systematic errors in the effective area and from deadtime
are comparable in magnitude to the statistical errors specified here.

The single detector count spectra obtained in ``spectroscopy mode'' were
processed as follows: the background was estimated from a spectrum of
140 seconds duration starting 202 seconds prior to the BATSE trigger.
Eight high range  (0.6--10.0 MeV) detector
spectra (4~s integration time each) covering a 32~s time interval
(Table~1), were background subtracted and summed.
For these data, the forward-folding fitting was done using  $\chi^2$
as the goodness-of-fit criterion.
The location of the burst (zenith distance = $56.4^\circ$)
results in an effective detector area of 541 cm$^2$, corresponding to
87\% of the on-axis area.
Assuming a single power law, the fit to the data from 1.0 to 8.5 MeV
(Fig.~2)
gives best fit parameter values of
normalization = ($1.49 \pm 0.10$)
photons cm$^{-2}$ s$^{-1}$ MeV$^{-1}$ at 1 MeV
and index $\beta$ = $-2.78 \pm 0.16$.  The fluence (1.0--8.5 MeV, 32~s) is
$7.93\times 10^{-5}$ erg cm$^{-2}$
We note a clear break of the data
below 1 MeV  where the
spectrum becomes flatter. Preliminary analysis of the low range
(0.3--1.5 MeV) spectroscopy data (not shown in Fig.~2)  covering
the same 32~s time interval also indicates a
spectral break at around 0.8 to 1 MeV.

\subsection{EGRET}

The Energetic Gamma-Ray Experiment Telescope (EGRET) has two methods of
detecting gamma-ray bursts.  The EGRET spark chambers are the primary
detector for the telescope and are sensitive to gamma rays above 30
MeV.  The spark chamber detector is turned on in response to an onboard
trigger provided by BATSE when a burst is detected within 40 degrees
of EGRET's principal axis. However, GRB~990123 was too
far off axis (56.4$^\circ$), so no information is available from the
spark chambers for this burst.

EGRET also detects gamma-ray bursts with the
Total Absorption Shower Counter (TASC), a $77 \times 77 \times 20$ cm
NaI(Tl) detector.  While acting principally as the calorimeter for the
spark chambers, it is also an independently triggered detector in the
energy range from 1 to 200 MeV.  The TASC is unshielded from charged
particles and has a high background rate, which is determined from the
time intervals just preceding and following the burst.  The response of
the TASC varies strongly with incident angle, because of the
intervening spacecraft material.  An EGS-4 Monte Carlo code is used
with the CGRO mass model to determine the response of the detector as a
function of energy and incident photon direction.  In normal mode a
spectrum is accumulated every 32.768 seconds.  Shorter time intervals
are also accumulated following a BATSE burst trigger; however, these
short intervals were prior to the peak of this burst since BATSE
triggered $\approx$ 25 seconds prior to the peak of the emission.

The burst was detected in the EGRET TASC in two
successive 32.768 second intervals.  The first is from $-$0.057  to
32.711~s relative to the BATSE trigger, the second from 32.735 to
65.503~s.
The break between the two records occurs roughly between the two major
spikes.  The differential photon spectrum is well fit by a power law,
$F(E) = k(E/1 {\rm MeV})^{\beta}$,
in each separate time interval as well as when
the intervals are combined.  The combined 65.5 second interval (Fig.~2) is fit
by $\beta = -2.71\pm 0.08$ and $k=1.09 \pm 0.07$
photons cm$^{-2}$ s$^{-1}$ MeV$^{-1}$.
The
normalization $k$ and the points in Fig.~2 are calculated assuming that
the excess flux above background was emitted
over 32.7~s, the time interval of the high energy emission from the
burst as seen in the light curves of BATSE and COMPTEL shown in
Fig.~1.

The separate spectral fits of the two major spikes show no evidence of
spectral evolution. The differential photon spectral index,  $\beta$,
for the first spectrum is $-2.56\pm 0.11$, and for the second is
$-2.78\pm 0.15$.  The normalization $k$ is $0.97 \pm 0.08$ for the first
spectrum and $1.20 \pm 0.13 $ for the second spectrum, where the time
intervals for the emissions of the first and second spectra are
assumed to be 20.4~s and 12.3~s, respectively.

\section{DISCUSSION}
\subsection{Optical}
The detection of simultaneous optical emission with ROTSE
is the primary distinction of GRB~990123, and the obvious question is
the relationship between the optical and gamma-ray fluxes. The optical
and gamma-ray
light curves are clearly very different: the optical flux rose by an
order of magnitude from the first to second observation, and then
fell by a factor of 2.5 to the third one. On the other hand, the
first optical observation occurred during the first hard gamma-ray
spike, while the second and third observations occurred during the
soft tail after the two main gamma-ray spikes. Indeed, the
20--2000~keV energy flux, calculated by integrating over fits to
the spectrum, is an order of magnitude higher during the first
optical observation than during the subsequent two, which had
comparable fluxes. Thus the optical and high-energy fluxes are not
directly proportional.

But perhaps the optical and gamma-ray emission are part of one
component, and the relative fluxes change as the spectrum evolves?
Figure~4 shows the observed optical and gamma-ray spectra for the
three observations.
The curves show the photon models obtained from fitting the BATSE data.
Note the extent of the X-ray excess on Figures 2, 3 and 4.
As can be seen, the gamma-ray spectra do not
extrapolate down to the optical band without an unseen break or
bend. We have included the maximum $3\sigma$ variations in the low-energy
gamma-ray spectrum permitted by the data; these
extrapolations underestimate the optical flux by an order of
magnitude or more. Perhaps the optical emission and the X-ray
excess observed during the first ROTSE observation are part of the
same component? We extrapolated a power law with index $\alpha_{ox}
=-0.243$ connecting the optical and 10~keV X-ray fluxes during this
first observation. Figure~4 shows that a similar power law
over predicts the X-ray flux by more than an order of magnitude
during the second and third optical observations.
We conclude that the optical emission is
in excess of the low-energy extrapolation of the burst spectrum,
which suggests that the optical and high-energy emission
originate from different shocks in the relativistic outflow.

\begin{figure*}[tbp!]
\begin{center}
\mbox{
\psfig{file=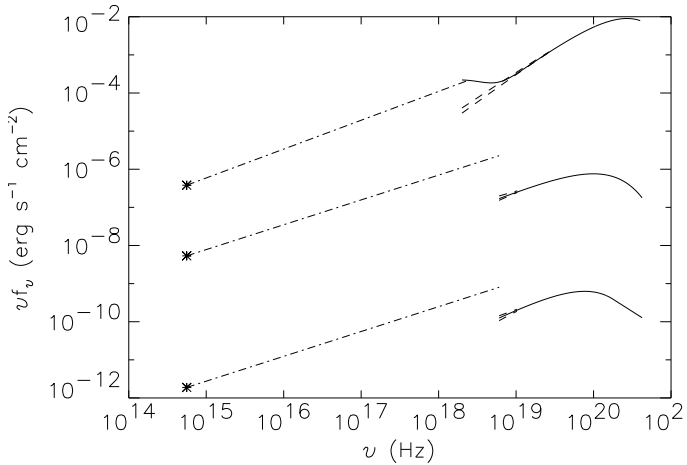,width=90mm,%
bbllx=64bp,bblly=350bp,bburx=540bp,bbury=700bp,clip=}
\hspace{4mm}
\begin{minipage}[b]{85mm}
\caption{Optical flux and gamma-ray spectra for the time intervals of
the 3 ROTSE observations.
The curves for the first (third) time
interval have been shifted up (down)  by a factor of 1000. The data
points ``*'' on the left are the ROTSE observations, while the solid
curves on the right are the best fit models to the BATSE SD
spectra. The dashed curves on the low-energy end of the gamma-ray
spectra show the spectra for $3\sigma$ variations in $\alpha$, the
GRB function's low-energy spectral index. The dot-dashed curve is
the extrapolation of the optical flux to 10~keV using
$\alpha_{ox}$, the power law index connecting the observed optical
and 10~keV X-ray fluxes for the first ROTSE time interval.}
\end{minipage}
}
\end{center}
\end{figure*}

\subsection{X-Ray Emission}

Also interesting is the observation that the X-ray excess and the gamma-ray
emission are apparently not related. That is, the excess occurs during the
two main spikes in the gamma-ray time history, but not in the time interval
afterwards, as well as being absent during a small flaring episode on the
falling portion
of the first main spike.
The X-ray time history in the 8--13 keV band shows
interesting behavior during the tail end of the burst: rather than the flat
shoulder present in the gamma-rays, there is a somewhat steady fall-off in
flux from T$+50$ s to T$+130$ s. At the same time, the low-energy power-law
index $\alpha$
makes a transition from $\sim -0.5$ to $\sim -1.0$ (see Fig.~1). Thus,
the portion of the burst associated with the X-ray excess is also associated
with a hard low-energy power-law index; when the low-energy index softens, the
excess disappears.
A possible reason is that the change in slope $\alpha$ causes the X-ray
extension of the
gamma-ray spectrum to mask the low-energy excess.
However, the flux predicted by the GRB function fit to the data of the
second ROTSE interval is inadequate to mask the X-ray excess observed
during the first ROTSE interval.
This suggests that the X-ray component has
disappeared by the second half of the burst---making GRB~990123 the first
burst in which BATSE has reported evolution in this component.
In that case, it is interesting
that the X-ray excess is not correlated with the optical emission, which peaks
during the second half of the burst.

There are several burst models that support the existence of excess X-rays
simultaneously with prompt burst emission. The most successful of these,
the Compton attenuation model (CAM---Brainerd 1994), predates the discovery
of the excess. In this model, a beam of photons is scattered
out of the line of sight by an optically-thick medium via
Compton scattering in the Klein-Nishina limit, which is less efficient at higher
photon energies than Thompson scattering. It is this change in efficiency that
modifies an assumed input power-law spectrum into the observed shape, with a
universal break energy in the rest frame determined by the physics.
The low-energy upturn is determined
by the density of the intervening scattering material. In the  simplest
form of the CAM, there is no allowance  for changes in the optical depth as a
function of time, yet the change in amplitude of the X-ray upturn in
the present burst seems to require this.
When we fit the BATSE spectral data for the first ROTSE interval with the CAM,
a $\chi^2$ value of
511 for 400 degrees of freedom is obtained.
Values of reduced $\chi^2$ above unity are typical of
very bright bursts like GRB~990123 because systematic deviations dominate
the statistical fluctuations.
The best-fit value for the red-shift is 1.19 $\pm 0.06$; while
$\chi^2$ increases by 27 to 538 when the red-shift is required to be 1.61.
When the GRB model is fit to all of the BATSE data, including the SD
discriminator points with their evidence for an X-ray excess, the same
$\chi^2$ value of 538 is obtained.
If the SD discriminator data are ignored, the GRB model has a substantially
better $\chi^2$ value than the CAM model, indicating the power of these data
for testing models with different X-ray predictions.

Other burst models have various levels of
compatibility with an excess of X-rays, but few require it. Usually, a
separate,
lower-energy spectral component would indicate that Compton up-scattering
is producing the observed gamma-rays from a seed population of X-rays.
One would expect the two components to be highly correlated in time,
which is not observed here. The input spectral shape is modified by the
distribution of energetic particles that are doing the up-scattering, so one
can
infer a power-law  index for the particles
from the photon continuum spectral index.
Finally, it is also not clear if there is any connection between the X-ray
flux
and the burst afterglow, presumably created in this case by
a shock formed when relativistically-expanding material from the burst meets
material external to the system, such as interstellar material, or the remnant
of a wind from the burst's progenitor (Sari \& Piran 1999a). Although the
component of the aftershock related to the optical flash (the reverse
shock) is
not expected to produce any detectable X-ray component, the forward shock is.
The observation that the hard X-rays (top lightcurve of Fig.~1),
have a maximum
after the occurrence of the two main spikes in the gamma-rays,
may indicate the peak of the forward shock in this band.

\subsection{Afterglow}

Park et al. (1997a) introduced three dimensionless statistics to
compare optical detections or upper limits with the observed
gamma-ray emission. Behind these statistics are implicit
assumptions about the relationship between the emission in these
two bands. The first two statistics can be calculated only if the
high-energy emission is still detectable at the time of the optical
emission. Comparing the optical and gamma-ray flux densities
$f_\nu$ (in units of erg s$^{-1}$ cm$^{-2}$ Hz$^{-1}$), is
$R_1=f_\nu(\nu_{\rm opt})/f_\nu(\nu_{\gamma})$, where $\nu_{\rm
opt}$ and $\nu_{\gamma}$ are fiducial optical and gamma-ray
frequencies (here the V band's $5.6\times10^{14}$~Hz and
100 KeV $=2.4\times10^{19}$~Hz).
Comparing the optical flux  to the gamma-ray  flux density
$f_\gamma$ (erg s$^{-1}$ cm$^{-2}$) is $R_2= f_\nu(\nu_{\rm opt})
\nu_{\rm opt}/f_\gamma$. Finally, $R_3= f_\nu(\nu_{\rm opt})
/\phi_\gamma$ compares the optical flux density to the gamma-ray
energy fluence $\phi_\gamma $ (erg cm$^{-2}$) from the beginning of
the burst until the time of the optical observation. $R_3$ is
relevant and calculable whether or not the high-energy emission is
detectable at the time of the optical observations; however the
time between the high-energy emission and the optical observation
is physically interesting. While these statistics do depend on the
fiducial optical and gamma-ray frequencies which are chosen because
of detector capabilities, they do not depend on instrumental
quantities such as the length of the optical exposure. Table~2
shows the values of these 3 statistics for the 6 ROTSE detections
of \g123. Also shown are typical upper limits obtained with the Gamma-Ray
Optical Counterpart Search Experiment (GROCSE), ROTSE's
less-sensitive predecessor, and the upper limits for GRB~970223
(Park et al. 1997b) obtained with the Livermore Optical Transient Imaging
System (LOTIS), an instrument very similar in design and operation
to ROTSE (LOTIS did not attempt to observe \g123\ because of bad
weather---H.-S.~Park, personal communication 1999). As can be seen,
the ROTSE detections are below the GROCSE upper limits, but are
comparable to the upper limits for GRB~970223. Only with more
detections and upper limits from ROTSE, LOTIS and similar
instruments will we determine whether \g123's optical emission was
typical.

\begin{deluxetable}{l r r r r r r r r}
\tablefontsize{\scriptsize}
\tablecolumns{9}
\tablewidth{0pc}
\tablecaption{Optical--Gamma-Ray Emission Comparison}
\tablehead{
\colhead{Obs.\tablenotemark{a}} &
\colhead{$t$\tablenotemark{b}} &
\colhead{$f_\nu (\nu_{\rm opt})$\tablenotemark{c}} &
\colhead{$f_E (E_{\rm 100keV})$\tablenotemark{d}} &
\colhead{$f_\gamma$\tablenotemark{e}} &
\colhead{$\phi_\gamma$\tablenotemark{f}} &
\colhead{$R_1$\tablenotemark{g}} &
\colhead{$R_2$\tablenotemark{h}} &
\colhead{$R_3$\tablenotemark{i}}
}
\startdata
\g123-1 &  22.2 & $6.79\times10^{-25}$ & $1.07\times10^{-8}$ &
$2.14\times10^{-5}$ & $7.78\times10^{-5}$ & $1.53\times10^{1}$ &
$1.78\times10^{-5}$ & $8.73\times10^{-21}$ \nl
\g123-2 &  47.4 & $9.55\times10^{-24}$ & $3.93\times10^{-9}$ &
$1.68\times10^{-6}$ & $3.05\times10^{-4}$ & $5.87\times10^{2}$ &
$3.19\times10^{-3}$ & $3.13\times10^{-20}$ \nl
\g123-3 &  72.7 & $3.37\times10^{-24}$ & $4.13\times10^{-9}$ &
$1.82\times10^{-6}$ & $3.39\times10^{-4}$ & $1.97\times10^{2}$ &
$1.04\times10^{-3}$ & $9.95\times10^{-21}$ \nl
\g123-4 & 281.4 & $1.87\times10^{-25}$ & --- & --- &
$2.65\times10^{-4}$ & --- & --- & $7.06\times10^{-22}$ \nl
\g123-5 & 446.7 & $9.12\times10^{-26}$ & --- & --- &
$2.65\times10^{-4}$ & --- & --- & $3.44\times10^{-22}$ \nl
\g123-6 & 612.0 & $5.60\times10^{-26}$ & --- & --- &
$2.65\times10^{-4}$ & --- & --- & $2.11\times10^{-22}$ \nl
GROCSE\tablenotemark{j} & $\sim25$ & --- & --- & --- & --- &
$<1.5\times10^{4}$ & $<4\times10^{-2}$ & $<3\times10^{-18}$ \nl
GRB~970223\tablenotemark{k} & 11 & $<1.4\times10^{-24}$ &
$1.4\times10^{-9}$ & $6.09\times10^{-7}$ & $4.8\times10^{-5}$ &
$<2.4\times10^{2}$ & $<1.3\times10^{-3}$ & $<2.9\times10^{-20}$
\nl
\enddata
\tablenotetext{a}{\footnotesize Optical observation.}
\tablenotetext{b}{\footnotesize Time in seconds from BATSE trigger to beginning of
optical observation.}
\tablenotetext{c}{\footnotesize V band optical flux density
(erg s$^{-1}$ cm$^{-2}$ Hz$^{-1}$).}
\tablenotetext{d}{\footnotesize 100~keV flux density
(erg s$^{-1}$ cm$^{-2}$ keV$^{-1}$).}
\tablenotetext{e}{\footnotesize 20--2000~keV gamma-ray energy flux
(erg s$^{-1}$ cm$^{-2}$).}
\tablenotetext{f}{\footnotesize 20--2000~keV energy fluence from the beginning 
of the burst to the middle of the optical observation (erg cm$^{-2}$).}
\tablenotetext{g}{\footnotesize Ratio of optical to gamma-ray flux densities,
$R_1=2.42\times10^{17} f_\nu (\nu_{\rm opt})/f_E (E_{\rm 100keV})$
(the constant is necessary for units conversion).}
\tablenotetext{h}{\footnotesize Ratio of optical flux density to gamma-ray
energy flux, $R_2=f_\nu (\nu_{\rm opt}) \nu_{\rm opt}/f_\gamma$.}
\tablenotetext{i}{\footnotesize Ratio of optical flux density to gamma-ray
energy fluence, $R_3=f_\nu (\nu_{\rm opt})/\phi_\gamma$.}
\tablenotetext{j}{\footnotesize Typical GROCSE upper limits from Park
et al. (1997a).}
\tablenotetext{k}{\footnotesize From Park et al. (1997b).}
\end{deluxetable}

The $R_3$ statistic can also compare the optical afterglows to the
gamma-ray emission; Table~3 shows $R_3$ for $R$ band afterglow
emission after 1~day for a number of recent bursts. These values of
$R_3$ are usually smaller than the detections and upper limits
during and immediately after the burst because it is possible to
observe the optical transient with large telescopes once the
position is known accurately. We would like to use $R_3$ values
calculated at the same time and optical frequency
in the rest frame of the burst.
Assuming that the afterglow flux has power law spectrum
and temporal decay, $f_\nu \propto \nu^\epsilon t^\delta$, then
$R_{3,\hbox{burst}}=R_{3,\hbox{observer}}(1+z)^{\delta-\epsilon}$.
For most afterglows $\epsilon\sim\delta\sim -1$ (e.g., Galama et
al. 1997, 1998, 1999; Bloom et al. 1998), and by afterglow theory
$\delta-\epsilon = (\delta+1)/3$ for adiabatic cooling; thus
$R_{3,\hbox{burst}}$ has at most a weak dependence on redshift. As
can be seen, there is a wide range of values, which confirms that
afterglows are not all alike. Indeed, the $R_3$ value and the
isotropic gamma-ray energy are inversely correlated, which results
from a broader energy distribution than the distribution of $R$
band specific luminosities in the bursts' frame (i.e., the optical
flux converted into a luminosity emitted by the afterglow). This
suggests that the efficiency with which the total energy is
converted into emission varies with wavelength band. For example,
in the standard model, the energy release by internal shocks
depends on details of the relativistic flow, which can vary from
burst to burst, while the kinetic energy of the flow is expended at
the external shock; thus the distribution of afterglow energies may
be narrower than that of the burst itself.

\begin{deluxetable}{lccccccl}
\tablefontsize{\scriptsize}
\tablecolumns{8}\footnotesize 
\tablewidth{0pc}
\tablecaption{Burst Energetics}
\tablehead{
\colhead{GRB} &
\colhead{$z$} &
\colhead{$\phi_\gamma$ ($>$ 20 keV)\tablenotemark{a}} &
\colhead{$E$\tablenotemark{b}} &
\colhead{$m_R$ (1 day)\tablenotemark{c}} &
\colhead{$f_\nu(\nu_R)$\tablenotemark{d}} &
\colhead{$R_3$\tablenotemark{e}} &
\colhead{Ref.\tablenotemark{f}}
}
\startdata
970228 & & $2.0\times 10^{-6}$ &
& 21.1$\pm$0.15 & 1.12$\times 10^{-5}$ & 5.6$\times 10^{-23}$ & 1,2
\nl
970508 & 0.835 & 3.7$\times 10^{-6}$ &
5.3$\times 10^{51}$ & 20.85$\pm$0.05 & 1.32$\times 10^{-5}$ &
3.8$\times 10^{-23}$ & 3,4,5 \nl
971214 & 3.418 & 1.1$\times 10^{-5}$ &
2.5$\times 10^{53}$ & 23.0$\pm$0.22 & 1.94$\times 10^{-6}$ &
1.8$\times 10^{-24}$ & 6,7 \nl
980326 & & $2.2\times 10^{-6}$ & &
23.39$\pm$0.12 & 1.36$\times 10^{-6}$ & 6.2$\times 10^{-24}$ & 8
\nl
980329 & $\sim$5 & 5.0$\times 10^{-5}$ &
2.4$\times 10^{54}$ & 23.9$\pm$0.2 & 8.48$\times 10^{-7}$ &
1.7$\times 10^{-25}$ & 9,10 \nl
980519 & & $1.1\times 10^{-7}$ & &
21.28$\pm$0.13 & 9.47$\times 10^{-6}$ & 8.6$\times 10^{-22}$ &
11,12
\nl
980613 & 1.096 & 1.7$\times 10^{-6}$ &
4.2$\times 10^{51}$ & 23.2$\pm$0.5 & 1.61$\times 10^{-6}$ &
9.5$\times 10^{-24}$ & 13,14 \nl
980703 & 0.966 & 4.6$\times 10^{-5}$ &
8.9$\times 10^{52}$ & 21.18$\pm$0.10 & 1.04$\times 10^{-5}$ &
2.3$\times 10^{-24}$ & 15,16 \nl
990123 & 1.61 & 3.0$\times 10^{-4}$ &
1.6$\times 10^{54}$ & 20.4$\pm$0.3 & 2.13$\times 10^{-5}$ &
7.1$\times 10^{-25}$ & this paper \nl
\enddata
\tablenotetext{a}{\footnotesize Energy fluence, in erg cm$^{-2}$.}
\tablenotetext{b}{\footnotesize Total energy in ergs if radiated
isotropically,
assuming $H_0 = 70$ km s$^{-1}$ Mpc$^{-1}$, $\Omega_0 = 0.3$, and
$\Lambda=0$.}
\tablenotetext{c}{\footnotesize R band magnitude after 1 day.}
\tablenotetext{d}{\footnotesize Equivalent R band flux density, in Jy.}
\tablenotetext{e}{\footnotesize Ratio of optical flux density to
gamma-ray energy
fluence, $R_3=f_\nu (\nu_{R})/\phi_\gamma$.}
\tablenotetext{f}{\footnotesize References: 1.~Galama et al. (1997);
2.~Fruchter et al.
(1999a); 3.~Galama et al. (1998); 4.~Metzger et al. (1997); 5.~Bloom
et al. (1998); 6.~ Diercks et al. (1997); 7.~Kulkarni et al.
(1998); 8.~Groot et al. (1998); 9.~Palazzi et al. (1998);
10.~Fruchter (1999); 11.~Vrba et al. (1998); 12.~Djorgovski et al.
(1998b); 13.~Djorgovski et al. (1998c); 14.~Djorgovski et al.
(1999); 15.~Vreeswijk et al. (1999); and 16.~Djorgovski et al.
(1998a).}
\end{deluxetable}

\subsection{Intensity}

\g123's fluence $\phi_\gamma=3.0\times10^{-4}$ erg cm$^{-2}$ places
it firmly on the $N(>\phi_\gamma) \propto \phi_\gamma^{-3/2}$
portion of the fluence distribution---only
0.4\% of the bursts observed by BATSE have had higher fluences.
By photon peak flux, only 2.5\% of BATSE bursts have been more intense.
Whatever the intensity
measure, peak photon flux $P$ or fluence $\phi_\gamma$, the high
end of BATSE's cumulative intensity distribution is a power law
with an index of $-$3/2. In the simplest cosmological model standard
candle bursts occur at a constant rate per comoving volume; bursts
in nearby Euclidean space produce the $\phi_\gamma^{-3/2}$ portion
of the cumulative peak flux distribution while the bend in this
distribution results from bursts at sufficiently high redshift
($z>0.3$) that spacetime deviates from Euclidean. Thus we would
expect \g123\ to occur at low redshift. However the absorption
lines in the optical transient's spectrum show that $z\ge1.61$ for
\g123\ (Kelson et al. 1999, Hjorth et al. 1999).
\g123\ demonstrates that
the $\phi_\gamma^{-3/2}$ portion of the peak flux distribution does
not originate only in nearby Euclidean space.
Consequently, the burst
rate per comoving volume must vary with redshift, with the burst
rate and cosmological geometry ``conspiring'' to produce the
Euclidean-like $\phi_\gamma^{-3/2}$ dependence. Indeed it has been
suggested that the burst rate is proportional to the star formation
rate (e.g., Wijers et al. 1998; Totani 1997), which has decreased
precipitously since the epoch corresponding to $z\sim 1.5$.

The redshift implies that
\g123's energy release was at least $1.6\times 10^{54}$ erg ($H_0=70$ km
s$^{-1}$ Mpc$^{-1}$, $\Omega=0.3$ and $\Lambda=0$), if the burst
radiates isotropically, comparable to the energy release in
GRB~980329, if $z\sim 5$ for this burst (Fruchter 1999).
Energy releases are given as isotropic equivalents, even though there
are indications of beaming
(Fruchter et al. 1999b, Kulkarni et al. 1999, but see M\'esz\'aros \& Rees 1999)
because the beaming angles are unknown.
See Table~3 for a list of the energies of the bursts with redshifts.
Both GRB~971214 ($z=3.418$---Kulkarni et al. 1998) and GRB~980703
($z=0.966$---Djorgovski et al. 1998a) had energies of
$\sim10^{53}$~erg while GRB~970508 ($z=0.835$---Metzger et al.
1997; Bloom et al. 1998) and GRB~980613 ($z=1.096$---Djorgovski et
al. 1999) had energies of $\sim6\times10^{51}$ erg. Thus we see
that the standard candle assumption behind the simplest
cosmological model is violated, and the intrinsic energy
distribution is very broad.

Thus two basic premises of the
simplest cosmological model are violated. The burst rate and the
energy release distribution must now be determined empirically,
with the peak flux distribution providing a constraint on these
distributions. Note that it may not be possible to separate the
burst rate into independent functions of redshift and energy.
\section{SUMMARY}
The high-energy emission of \g123\ was fairly typical of gamma-ray
bursts.
The spectrum above 25~keV can be fit satisfactorily by the
simple four-parameter GRB function.
The parameter $E_{\rm p}$, the energy at which $\nu f_\nu$ peaks, reached
the  unusually high value $1470 \pm 110$ keV.
The typical spectral evolution patterns of
hardness-intensity correlation and hard-to-soft evolution
are observed.
An X-ray excess
is present below $\sim 15$~keV during the early part of the burst,
particularly during the first ROTSE observation. The burst's peak
flux and energy fluence place the burst in the top 2.5\% and 0.4\% of all
bursts observed by BATSE; these values also are clearly in the portion
of the intensity distribution which appears consistent with the
$-$3/2 slope power law  expected
for sources uniformly distributed in Euclidean space. However,
\g123's redshift of $z\ge1.61$ shows that the burst originated far
beyond nearby Euclidean space. This redshift and the fluence imply
an energy release of $\ge 1.6\times10^{54}$ ergs ($H_0=70$ km s$^{-1}$
Mpc$^{-1}$, $\Omega=0.3$ and $\Lambda=0$), if radiated
isotropically.
GRB~980329, if at $z \sim 5$ (Fruchter 1999),
has a similar energy release; the
determination of the redshift
of GRB~990123 via spectroscopic lines in the afterglow leaves no doubt
that isotropic-equivalent energy releases above $10^{54}$ ergs occur.

The gamma-ray and ROTSE optical fluxes are not correlated,
and the gamma-ray spectrum does not extrapolate down to the optical
observations. Thus there is no indication that the emissions in
these two bands are from the same component. Indeed, the standard
burst theory attributes the high-energy emissions to ``internal
shocks'' within a relativistic outflow, and the simultaneous
optical emission to a reverse shock that forms when the
relativistic outflow plows into the surrounding medium.

Various measures that relate the optical and gamma-ray emissions
show that the ROTSE detection of optical emission from \g123\ is
consistent with the upper limits on simultaneous optical emission
from previous bursts.  Therefore, only future optical observations
by ROTSE, LOTIS and similar robotic systems will determine whether
\g123\ was typical.

\acknowledgments

MSB and RDP acknowledge support from NASA grant NAG5-7927.
The gamma-ray burst research of DLB is supported by the {\it CGRO}
guest investigator program.
The COMPTEL project is supported in part through NASA grant
NAS~5-26646, DARA grant 50 QV 90968, and the Netherlands Organization
for Scientific Research (NWO).  AC is supported in part through the
hospitality of Wellesley College and NASA grant NAG5-7984.  MM is supported
in part through NASA grant NAG5-7829.
RMK and RDP acknowledge support through NASA grant NAG5-6747.
SMM acknowledges support from NASA grant NAG5-7894.


\clearpage

\begin{thebibliography}{}
%
\bibitem[a]{b}Akerlof, C. W., et al. 1999a, GCN \#205
%
\bibitem[a2]{b12}------------ 1999b, Nature, in press
%
%
\bibitem[a42]{b232}Band, D., et al. 1993, ApJ, 413, 281
%
\bibitem[casd]{dffd}Bloom, J. S., Djorgovski,~S.~G., Kulkarni,~S.~R., \&
Frail,~D.~A. 1998, ApJ, 507, L25
%
%
\bibitem[dd]{what0}Brainerd, J. J. 1994, ApJ, 428, 21
%
\bibitem[brev]{what1}Briggs, M. S. 1996, in AIP Conf. Proc. 384,
ed. C. Kouveliotou, M. Briggs \& G. Fishman, 133
%
\bibitem[csc]{csc}Burenin, R. A., et al. 1999, A\&A, submitted
[astro-ph/9902006]
%
\bibitem[aza]{bzb}Connors, A., \& Hueter, G. J. 1998, ApJ, 501,
307
%
\bibitem[fd2]{d44}Costa, E., et al. 1997, Nature, 387, 783
%
\bibitem[kdskho]{yhtyh}Diercks, A., Deutsch, E. W., Wyse,~R.,
Gilmore,~G., Corson,~C., Castander,~F.~J., \& Turner,~E. 1997, IAU
Circ. 6791
%
\bibitem[gasdf]{hagh}Djorgovski, S. G., Kulkarni, S. R., Bloom, J.
S., Goodrich,~R., Frail,~D.~A., Piro,~L., \& Palazzi,~E. 1998a,
ApJ, 508, L17
%
\bibitem[g1as5df]{h1a3gh}------------ 1998b, GCN \#79
%
\bibitem[g14asdf]{h16agh}------------ 1998c, GCN \#117
%
\bibitem[g1asdf]{h1agh}------------ 1999, GCN \#189
%
\bibitem[aa]{bb}Feroci, M., et al. 1999, IAU Circ. \#7095
%
\bibitem[a233]{b223}Ford, L., et al. 1995, ApJ, 439, 307
%
\bibitem[a1a]{b1b}Frail, D. A., et al. 1999, GCN \#211
%
\bibitem[851gv]{121yf}Fruchter, A. 1999, ApJ, in press
[astro-ph/9810224]
%
\bibitem[85gv]{12yf}Fruchter, A., et al. 1999a, ApJ, in press
[astro-ph/9807295]
%
\bibitem[xyz11]{what11}Fruchter, A., et al. 1999b, submitted
[astro-ph/9902236]
%
\bibitem[aSD~1fa]{413t}Galama, T., et al. 1997, Nature, 387, 479
%
\bibitem[asdfa]{43t}------------ 1998, ApJ, 497, L13
%
\bibitem[asdf2a]{432t}------------ 1999, Nature, in press [astro-ph/9903021]
%
\bibitem[giblin]{what3}Giblin, T., et al., 1999, ApJ, submitted
%
\bibitem[g1a54sdf]{h1a23gh}Groot, P. J., et al. 1998, ApJ, 502,
L123
%
\bibitem[acc]{bccb}Heise, J., et al. 1999, IAU Circ. \# 7099
%
\bibitem[ac]{bcb}Hjorth, J., et al. 1999, GCN \#219
%
\bibitem[a22]{a22}in't Zand, J. J. M., et al. 1998, ApJ, 505, L119
%
\bibitem[cc]{dd}Kelson, D. D., Illingworth, G. D., Franx, M.,
Magee,~D., van~Dokkum,~P.~G. 1999, IAU Circ. \#7096
%
\bibitem[xyz1]{what4}
Kippen, R. M., Ryan, J. M., Connors, A., Winkler, C., Sch\"onfelder,
V., Kuiper, L., McConnell, M., Varendorff, M., Hermsen, W., and
Collmar, W. 1998, Adv. Space Res., 22, No. 7, 1097
%
\bibitem[ee]{fe}Kippen, R.~M., et al. 1999, GCN \#224
%
\bibitem[er]{re}Kulkarni, S., et al. 1998, Nature, 393, 35
%
\bibitem[xvy12]{what12}Kulkarni, S. R., et al. 1999, Nature, in press
[astro-ph/9902272]
%
\bibitem[xyz5]{what5}
Loredo, T. J., 1992, in Statistical Challenges in Modern Astronomy,
ed. G. J. Babu \& E. D Fiegelson (New York: Springer-Verlag)
%
\bibitem[xyz6]{what6}
Matz, S., et al. 1999, ApJ,  in preparation
%
\bibitem[gs]{gas}M\'esz\'aros, P., \& Rees, M. J. 1997, ApJ, 476,
232
%
\bibitem[xyz13]{what13}M\'esz\'aro, P., \& Rees, M. J., 1999, submitted
[astro-ph/9902367]
%
\bibitem[gg]{hg}Metzger, M., et al. 1997, Nature, 387, 878
%
\bibitem[g]{h}Odewahn, S. C., et al. 1999, GCN \#201
%
\bibitem[hhfghf]{trhjd}Palazzi, E., et al. 1998, A\&A, 336, L95
%
\bibitem[a2]{b2}Park,~H.-S., et al. 1997a, ApJ, 490, 99
%
\bibitem[a3]{b3}------------ 1997b, ApJ, 490, L21
%
\bibitem[priran]{what2}
Piran, 1999, Physics Reports, in press [astro-ph/9810256]
%
\bibitem[e]{f}Piro, L., et al. 1999, GCN \#199, \#202, \#203
%
\bibitem[adg2]{bfh2}Preece, R., et al. 1996, 473, 310
%
\bibitem[xzy21]{xyz21}Preece, R., et al. 1999, ApJ, in preparation
%
\bibitem[xyz20]{what20}Rees, M. J., \& M\'esz\'aros, P. 1992, MNRAS, 258, 41
%
\bibitem[f7]{k778}Sari, R., \& Piran, T. 1999a, ApJ,
submitted [astro-ph/9901338]
%
\bibitem[xyz10]{what10}Sari, R., \& Piran, T. 1999b, submitted
[astro-ph/9902009]
%
\bibitem[as3d]{yq3}Sazonov, S. Y., Sunyaev, R. A.,
Terekhov,~O.~V., Lund,~N., Brandt,~S., \& Castro-Tirado,~A.~J.
1998, A\&ASuppl., 129, 1
%
\bibitem[xyz7]{what7}
Sch\"onfelder, V. et al., 1993, ApJS, 86, 629
%
\bibitem[fs]{sd}Totani, T. 1997, ApJ, 486, L71
%
\bibitem[hjktyuty]{12tggdf}Vrba, F. J., et al. 1998, GCN \#83
%
\bibitem[hgf]{gfg}Vreeswijk, P. M., et al. 1999, ApJ, in press
%
\bibitem[ga]{ga}Wijers, R. A. M. J., et al. 1998, MNRAS, 294, L13
%
\bibitem[asd]{asdf}Yoshida, A., et al. 1989, PASJ, 41, 509
%
%
\end{thebibliography}
\end{document}